\documentclass[useAMS,a4,usenatbib]{mn2e}
\input psfig.sty
\usepackage{graphicx}
\usepackage{wasysym}

\usepackage{placeins}

\def \xoff {\ifmmode x_{\rm off} \else $x_{\rm off}$ \fi}
\def \rhorms {\ifmmode \rho_{\rm rms} \else $\rho_{\rm rms}$ \fi}
\def \rhocrit {\ifmmode \rho_{\rm crit} \else $\rho_{\rm crit}$ \fi}

\def \etal {et~al.~}
\def \chisq  {\ifmmode  \chi^2   \else  $\chi^2$  \fi}  
\def \chisqr {\ifmmode \chi^2_{\rm r} \else $\chi^2_{\rm r}$ \fi}
\def \spose#1{\hbox  to 0pt{#1\hss}}  
\def \lta{\mathrel{\spose{\lower 3pt\hbox{$\sim$}}\raise  2.0pt\hbox{$<$}}}
\def \gta{\mathrel{\spose{\lower  3pt\hbox{$\sim$}}\raise 2.0pt\hbox{$>$}}}
 
\def \ha  {\ifmmode H\alpha \else H$\alpha $ \fi}

\def \kms {\ifmmode  \,\rm km\,s^{-1} \else $\,\rm km\,s^{-1}  $ \fi }
\def \kpc {\ifmmode  {\rm kpc}  \else ${\rm  kpc}$ \fi  }  
\def \Msun {\ifmmode M_{\odot} \else $M_{\odot}$ \fi} 
\def \hMsun {\ifmmode h^{-1}\,\rm M_{\odot} \else $h^{-1}\,\rm M_{\odot}$ \fi}
\def \hhMsun {\ifmmode h^{-2}\,\rm M_{\odot}\else $h^{-2}\,\rm M_{\odot}$ \fi}
\def \Lsun {\ifmmode L_{\odot} \else $L_{\odot}$ \fi} 
\def \hhLsun {\ifmmode h^{-2}\,\rm L_{\odot} \else $h^{-2}\,\rm L_{\odot}$ \fi}

\def \LCDM {\ifmmode \Lambda{\rm CDM} \else $\Lambda{\rm CDM}$ \fi}
\def \sig8 {\ifmmode \sigma_8 \else $\sigma_8$ \fi} 
\def \OmegaM {\ifmmode \Omega_{\rm M} \else $\Omega_{\rm M}$ \fi} 
\def \OmegaL {\ifmmode \Omega_{\rm \Lambda} \else $\Omega_{\rm \Lambda}$\fi} 
\def \Deltavir {\ifmmode \Delta_{\rm vir} \else $\Delta_{\rm vir}$ \fi}

\def \rs {\ifmmode r_{\rm s} \else $r_{\rm s}$ \fi} 
\def \rrm2 {\ifmmode r_{-2} \else $r_{-2}$ \fi} 
\def \ccm2 {\ifmmode c_{-2} \else$c_{-2}$ \fi} 
\def \cvir {\ifmmode c_{\rm vir} \else $c_{\rm vir}$ \fi} 
\def \cbar {\ifmmode \overline{c} \else $\overline{c}$ \fi}

\def \R200 {\ifmmode R_{200} \else $R_{200}$ \fi} 
\def \Rvir {\ifmmode R_{\rm vir} \else $R_{\rm vir}$ \fi}

\def \v200 {\ifmmode V_{200} \else $V_{200}$ \fi} 
\def \Vvir {\ifmmode V_{\rm  vir} \else  $V_{\rm vir}$  \fi} 
\def  \Vhalo  {\ifmmode V_{\rm halo} \else $V_{\rm halo}$ \fi}

\def \M200 {\ifmmode M_{200} \else $M_{200}$ \fi} 
\def \Mvir {\ifmmode M_{\rm  vir} \else $M_{\rm  vir}$ \fi}  
\def \Mshell  {\ifmmode M_{\rm shell} \else $M_{\rm shell}$ \fi}

\def \Nvir {\ifmmode N_{\rm  vir} \else $N_{\rm  vir}$ \fi}  

\def \Jvir {\ifmmode J_{\rm vir} \else $J_{\rm vir}$ \fi} 
\def \Jshell {\ifmmode J_{\rm shell} \else $J_{\rm shell}$ \fi}

\def \Evir {\ifmmode E_{\rm vir} \else $E_{\rm vir}$ \fi} 

\def \lam {\ifmmode \lambda  \else $\lambda$ \fi} 
\def \lamp {\ifmmode \lambda^{\prime} \else $\lambda^{\prime}$  \fi} 
\def \lampc {\ifmmode \lambda^{\prime}_{\rm c} \else
  $\lambda^{\prime}_{\rm c}$  \fi} 
\def \lambar {\ifmmode \bar{\lambda}  \else  $\bar{\lambda}$  \fi}  
\def  \lampbar  {\ifmmode \bar{\lambda^{\prime}} \else
  $\bar{\lambda^{\prime}}$\fi} 
\def \siglam {\ifmmode \sigma_{\lambda} \else $\sigma_{\lambda}$ \fi} 
\def \siglamp {\ifmmode                \sigma_{\lambda^{\prime}} \else
$\sigma_{\lambda^{\prime}}$\fi}

\def \Rd {\ifmmode R_{\rm d} \else $R_{\rm d}$ \fi} 
\def \Rs {\ifmmode R_{\rm s} \else $R_{\rm s}$ \fi}  
\def \Rd {\ifmmode R_{\rm d} \else $R_{\rm d}$ \fi}  
\def \Rcool  {\ifmmode R_{\rm  cool}  \else $R_{\rm cool}$ \fi} 
\def \RIII {\ifmmode  3.2\Rs \else $3.2\Rs$ \fi} 
\def \RII {\ifmmode 2.2\Rs \else $2.2\Rs$  \fi} 
\def \Reff {\ifmmode R_{\rm eff} \else $R_{\rm  eff}$ \fi} 
\def  \rb {\ifmmode r_{\rm b}  \else $r_{\rm b}$ \fi}

\def  \Sigmacrit   {\ifmmode  \Sigma_{\rm  crit}   
\else  $\Sigma_{\rm crit}$\fi} 
\def \Sig0 {\ifmmode \Sigma_{0} \else $\Sigma_{0}$ \fi}

\def \muI {\ifmmode \mu_{0,I} \else $\mu_{0,I}$ \fi}

\def \mgal {\ifmmode m_{\rm gal} \else $m_{\rm gal}$ \fi} 
\def \md {\ifmmode m_{\rm d} \else $m_{\rm d}$ \fi} 
\def \ms {\ifmmode m_{\rm   s}   \else   $m_{\rm   s}$   \fi}   
\def   \mdbar   {\ifmmode {\overline{m}}_{\rm d} \else
  ${\overline{m}}_{\rm d}$ \fi} 
\def \msbar {\ifmmode  \bar{m}_{\rm  s}  \else  $\bar{m}_{\rm s}$
  \fi}  
\def  \Md {\ifmmode M_{\rm d}  \else $M_{\rm d}$ \fi} 
\def  \Ms {\ifmmode M_{\rm s} \else $M_{\rm  s}$ \fi} 
\def \Mb {\ifmmode  M_{\rm b} \else $M_{\rm b}$ \fi} 
\def \Mstar {\ifmmode  M_{\rm star} \else $M_{\rm star}$ \fi}
\def \Mdisc {\ifmmode M_{\rm disc} \else $M_{\rm disc}$ \fi}

\def \Jd {\ifmmode J_{\rm d} \else $J_{\rm d}$ \fi} 
\def \Jb {\ifmmode J_{\rm b} \else $J_{\rm b}$ \fi}  
\def \fb {\ifmmode  f_{\rm b} \else $f_{\rm b}$ \fi}

\def  \jd  {\ifmmode j_{\rm  d}  \else  $j_{\rm  d}$ \fi}  
\def  \jdmd {\ifmmode \frac{j_{\rm  d}}{m_{\rm d}} \else
  $\frac{j_{\rm d}}{m_{\rm d}}$ \fi} 
\def \fj {\ifmmode f_{\rm j} \else $f_{\rm j}$ \fi} 
\def \ft {\ifmmode f_{\rm t}  \else $f_{\rm t}$ \fi} 
\def  \fM {\ifmmode f_{\rm M} \else $f_{\rm M}$ \fi}

\def  \Vd {\ifmmode  V_{\rm  d}  \else $V_{\rm  d}$  \fi} 
\def  \Vcool {\ifmmode V_{\rm cool} \else $V_{\rm cool}$ \fi} 
\def \Vcirc {\ifmmode V_{\rm circ}  \else $V_{\rm circ}$  \fi} 
\def \VIII  {\ifmmode V_{3.2} \else $V_{3.2}$ \fi} 
\def  \VII {\ifmmode V_{2.2} \else $V_{2.2}$ \fi}
\def \Vobs {\ifmmode V_{\rm obs}  \else $V_{\rm obs}$ \fi} 
\def \Vdisc {\ifmmode V_{\rm disc} \else  $V_{\rm disc}$ \fi} 
\def \Vmax {\ifmmode V_{\rm  max} \else  $V_{\rm max}$  \fi} 
\def  \Vmaxobs{\ifmmode V_{\rm max}^{\rm obs}\else  $V_{\rm max}^{\rm
    obs}$\fi}  
\def \Vtot {\ifmmode V_{\rm tot} \else $V_{\rm tot}$  \fi} 
\def \Vrot {\ifmmode V_{\rm rot} \else  $V_{\rm rot}$  \fi} 
\def  \Vflat {\ifmmode  V_{\rm  flat} \else $V_{\rm flat}$ \fi}

\def \Ups {\ifmmode \Upsilon  \else $\Upsilon$ \fi} 
\def \YB {\ifmmode \Upsilon_B \else $\Upsilon_B$ \fi} 
\def \YI {\ifmmode  \Upsilon_I  \else $\Upsilon_I$ \fi} 
\def \DeltaIMF {\ifmmode \Delta_{\rm IMF} \else $\Delta_{\rm IMF}$ \fi}

\def\LCDM{$\Lambda$CDM }

\def\c200{$c_{200}$}

\title[Satellites in hydro and Nbody simulations] {Comparing galactic
  satellite properties in hydrodynamical and Nbody simulations}
\author[J.A. Schewtschenko \& A.V.  Macci\`o ] {Jascha
  A. Schewtschenko$^{1,2}$\thanks{jschewts@mpia.de}, Andrea
  V. Macci\`o$^{1}$\thanks{maccio@mpia.de} \\
  $^1$Max-Planck-Institut f\"ur Astronomie, K\"onigstuhl 17, 69117 Heidelberg, Germany\\
  $^2$Fakult\"at f\"ur Physik, Universit\"at Bielefeld, Postfach
  100131, 33501 Bielefeld, Germany}

\begin{document}
             
\date{submitted to MNRAS}
             
\maketitle           

\label{firstpage}
             
\begin{abstract}

  In this work, we examine the different properties of galactic satellites in
  hydrodynamical and pure dark matter simulations. We use three pairs of
  simulations (collisional and collision-less) starting from identical initial
  conditions. We concentrate our analysis on pairs of satellites in the hydro
  and Nbody runs that form from the same Lagrangian region.  We look at the
  radial positions, mass loss as a function of time and orbital parameters of
  these ``twin'' satellites.  We confirm an overall higher radial density of
  satellites in the hydrodynamical runs, but find that trends in the mass loss
  and radial position of these satellites in the inner and outer region of the
  parent halo differ from the pure dark matter case. In the outskirts of the
  halo ($\approx 70\%$ of the virial radius) satellites experience a stronger
  mass loss and higher dynamical friction in pure dark matter runs. The
  situation is reversed in the central region of the halo, where
  hydrodynamical satellites have smaller apocenter distances and suffer higher
  mass stripping. We partially ascribe this bimodal behaviour to the delayed
  infall time for hydro satellites, which on average cross the virial radius
  of the parent halo 0.7 Gyrs after their dark matter twins. Finally, we
  briefly discuss the implications of the different set of satellite orbital
  parameters and mass loss rates in hydrodynamical simulations within the
  context of thin discs heating and destruction.

\end{abstract}

\begin{keywords}
galaxies: haloes -- cosmology:theory, dark matter, gravitation --
methods: numerical, N-body simulation
\end{keywords}

\setcounter{footnote}{1}

\section{Introduction}
\label{sec:intro}

In the current paradigm of structure formation, large objects, such as
galaxies or clusters, are believed to form hierarchically, through a
'bottom-up' \citep{white1978} process of merging.
About a decade ago, N-body simulations attained sufficient dynamic
range to reveal that, in Cold Dark Matter (CDM) models, all haloes should
contain a large number of embedded subhaloes that survive the collapse
and virialization of the parent structure \citep{klypin1999,moore1999}.

The properties of subhaloes on different scales has been the subject
of many recent studies that have pushed the resolution of
dissipationless simulations \citep[e.g.][]{springel2001,deLucia2004,kravtsov2004,gao2004,
reed2005,diemand2007,zentner2005,springel2008}.  The kinetic properties of subhaloes are now
well understood - they make up a fraction of between 5 and 10\% of the
mass of virialized haloes, on scales relevant to observational
cosmology.

Most of these previous studies used dissipationless cosmological
simulations; although non-baryonic dark matter exceeds baryonic matter
by a factor of $\Omega_{dm}/\Omega_b \simeq 6$ on average \citep[e.g.][]{komatsu2009}
, the gravitational field in the central region
of galaxies is dominated by stars and gas.  The cooling baryons
increase the density in the central halo region mainly because of the
extra mass associated with the inflow, but also because of the
adiabatic contraction of the total mass distribution
\citep[e.g.][]{gnedin2004}. Since this process is active for {\it both} the host
halo and its subhaloes, it might be expected that subhaloes formed
within hydrodynamical simulations (including gas and stars) will
experience a different tidal force field and will themselves be
more robust to tidal effects.

Recently, a number of authors have examined the impact of baryonic
physics (gas cooling, star formation and feedback) on both the central
object and the satellite population in galaxy and cluster-sized haloes
\citep[e.g.][]{bailin2005,nagai2005,maccio2006,weinberg2008,romano2009,romano2010,libeskind2010,
sommer2010,duffy2010}.
\cite{maccio2006} simulated a Galactic mass halo twice - once considering pure DM
and once including baryons modeled with smoothed particles hydrodynamics, stopping the gas cooling at $z=1.5$. They found that the hydro run
produced an overabundance of subhaloes in the inner regions of the
halo as well as an increase by a factor of 2 in the absolute number of
subhaloes with respect to the DM run. \\
\indent The issue of the distribution and properties of 
galactic satellites in hydro and DM
simulations has been more recently revisited by
\cite{libeskind2010} and \cite{romano2010}.  \cite{libeskind2010} found results very similar to the work of
\cite{maccio2006}, with subhaloes in the hydro simulation 
being more radially concentrated than their dark
matter counterparts.  
They ascribe this effect to the higher central density of
subhaloes in hydro simulation (due to the collapse of baryons into
stars in the central region) that makes them more resilient to tidal
forces.  The increased mass in hydrodynamic subhaloes with respect to
dark matter ones causes dynamical friction to be more effective,
dragging the subhalo towards the centre of the host.

The overall properties of the satellite population in the study of
\cite{romano2010} are also consistent with \cite{maccio2006} and
\cite{libeskind2010}.  But, perhaps counter-intuitively, they find
that satellites in the hydro run are depleted at a faster rate than
the pure DM one within the central 30 kpc of the prime halo. According
to their analysis, \cite{romano2010} suggest that although the baryons
provide a substantial glue to the subhaloes, the main halo exhibits
the same trend.  This would assure a more efficient tidal disruption
of the hydro subhalo population 
in the inner region of the halo ($\approx 0.1$ of the virial radius).

Studies concerning the different
results of pure dark matter and hydrodynamical simulations are especially compelling within
the context of satellite effects on disk stability.
Several theoretical and numerical studies have been devoted to
quantifying the resilience of galactic disks to infalling satellites
\citep[e.g.][]{toth1992,quinn1993,velazquez1999,font2001,read2008,villalobos2008,moster2010}.
Recently \cite{kazantzidis2009} performed a
very detailed study of the dynamical response of thin
galactic disks to bombardment by cold dark matter substructure. They
used pure Nbody simulations of the formation of a Milky Way-like dark
matter halo to derive the properties of substructures and subsequently as initial conditions in subsequent high resolution
satellites-disk merger simulations.
Clearly, understanding {\it if} and {\it how} different possible orbital parameters 
and mass loss rates expected for satellites in hydro runs modify 
results previously obtained using pure Nbody simulations is of importance.

In this work we revisit the issue of the effects of baryonic physics on the
satellite population in galactic-size dark matter haloes. 
We improve the original study by \cite{maccio2006} in several aspects, namely  
with better parametrization of the baryonic physics, a full hydrodynamical approach 
down to redshift zero and a more extensive analysis of the satellite properties, including
the time evolution of mass loss, radial position, and orbital parameters (peri and apo-centre distances).
We start hydro and pure dark matter simulations from the same initial conditions and we focus our
analysis on ''twins'' satellites, i.e. (sub)structures that 
are formed from the same Lagrangian region for the initial conditions, which should therefore share
the same formation history in both simulation types.

The remainder of the paper is organized as follows: in
Section~\ref{sec:sims} we describe our simulations, and provide a brief summary of the numerical codes we use, including the technique employed to match satellites in the
different runs.  In Section~\ref{sec:res} we present our main
results, focusing on several satellite properties like radial position, orbital parameters, 
mass loss. Finally in Section~\ref{sec:conc} we summarize
and discuss our results.

\section{Numerical Simulations} 
\label{sec:sims}

The hydro-simulations were performed with {\sc gasoline}, a
multi-stepping, parallel Tree\-SPH $N$-body code \citep{Wadsley2004}.
We include radiative and Compton cooling for a primordial mixture of
hydrogen and helium.
The star formation algorithm is based on a Jeans instability criteria
\citep{Katz1992}, but simplified so that gas particles satisfying
constant density and temperature thresholds in convergent flows spawn
star particles at a rate proportional to the local dynamical time
\citep[see][]{Stinson2006}. The star formation efficiency was set to $0.05$
based on simulations of the Milky Way satisfying the \cite{kennicutt1998} Schmidt Law. The code also includes supernova feedback in the manner of \cite{Stinson2006}, and a UV background following
\cite{Haardt1996} (see \citep{Governato2007} for a more detailed
description of the code).

Three candidate haloes with masses similar to the mass of
our Galaxy ($M \sim 10^{12} \Msun$) were selected from an existing low resolution
dark matter simulation (300$^3$ particles within 90 Mpc) and subsequently
re-simulated at higher resolution.  These high resolution runs
are 8$^3$ times more resolved in mass than the initial set and we
included a gaseous component within the entire high resolution region.
Masses of the dark matter and gaseous particles are 
$m_{d} = 1.17 \times 10^6 \hMsun$ and $m_g = 2.3 \times 10^5 \hMsun$, respectively.
The dark matter has a spline gravitational softening length of 500
$h^{-1}$ pc and each component (dark and gas) consists of about $10^6$ particles  
in the high resolution region.  A list of galaxies properties can be found in Table \ref{table:haloes}.  A more detailed description of these hydro simulations (with particular attention to the
properties of the central galaxy) will appear in a forthcoming paper (Hernandez
\etal in prep).

\begin{table}
 \centering
 \begin{minipage}{140mm}
  \caption{Galaxies parameters}
  \begin{tabular}{lcccc}
\hline  Galaxy &  Mass  &  $R_{vir}$ & $N_{sat}$  \\
 &($10^{11}\hMsun)$&  (kpc/h)  & $(<R_{vir})$ \\
\hline 
G0 (DM/Hydro)  & 7.8/7.4  & 191/188 & 63/77 \\
G1 (DM/Hydro)  & 9.2/8.9  & 201/199 & 70/87 \\
G2 (DM/Hydro)  & 10.0/9.3 & 207/202 & 102/110 \\
\hline 
\label{table:haloes}
\end{tabular}
\end{minipage}
\end{table}

To run the dark matter only counter-parts of G0-G2 we use the Nbody
code {\sc pkdgrav} \citep{stadel2001}. This code is intimately related
with {\sc gasoline} which is its hydrodynamical extension. This
implies that DM particles are treated in exactly the same way in the
two codes (e.g. using the same numerical algorithms), allowing a
straight, direct comparison of the results.  The initial conditions
for the DM simulations are identical to those used for the hydro runs;
we simply transform all gas particles into dark matter particles. This
insures us both that the normalization and the phases of the initial
density perturbations are identical in the two runs.  These Nbody
simulations are the same presented in \cite{maccio2010}.

\begin{figure}
\psfig{figure=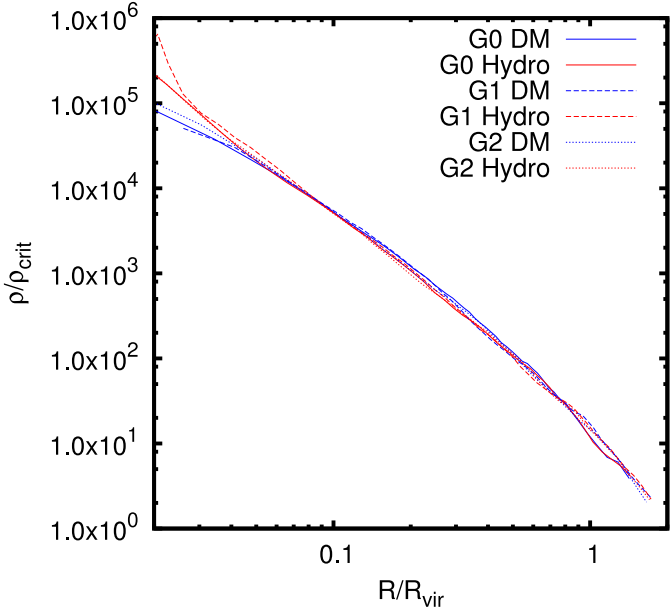,width=0.5\textwidth,angle=270}
\caption{Radial density profile for G0-G2 in the hydro (red lines) and
  DM (blue lines) simulations.}
\label{fig:densprof}
\end{figure}

Figure \ref{fig:densprof} shows the density radial density profiles of the G0-G2
galaxies in the two runs: hydro (red lines) and DM (blue lines).
As expected the profiles only diverge in the central regions due to the presence
of a baryonic core in the hydro runs.

\subsection{Halo finder and merger tree construction}

To identify subhaloes in our simulation we use the
MPI+OpenMP hybrid {\sc ahf} halo finder, available freely at http://www.popia.ft.uam.es/AMIGA) and described
in detail in \cite{Knollmann2009}. {\sc ahf} identifies local
over-densities in an adaptively smoothed density field as prospective
halo centers.  The local potential minima are computed for each of
these density peaks and the gravitationally bound particles are
determined. Only peaks with at least 50 bound particles are considered
to be haloes and retained for further analysis.  As subhaloes are embedded
within their respective host halo, their own density profile usually
shows a characteristic upturn at a radius $r_t \lta r_{\rm vir}$,
where $r_{\rm vir}$ is the actual (virial) radius of the satellites, were they found
in isolation.  We use this ``truncation radius'' $r_t$ as the
outer edge of the subhaloes and calculate all (sub-)halo properties (i.e. mass)
using only the gravitationally bound particles inside $r_t$.

As second step we build the merger trees for our galaxies
(both hydro and DM), in order to establish the dynamical history of
each infalling satellite across cosmic time.  For the purpose of
constructing an accurate tracking of each simulated (sub)halo, we
analyse 40 simulation outputs between $z=2.5$ and $z=0$.  We start from the
halofinder results at $z=2.5$ and follow these 
(sub)haloes through cosmic time, adding to the tracking procedure at each snapshot all new
haloes that cross our mass threshold.
For halo tracking, only dark matter particles are used.
We consider two criteria to decide if, when comparing two consecutive snapshots, 
halo 1 at one output time is the ``progenitor'' of halo 2 at the subsequent
time step: i) more than 50\% of the particles in halo
1 that end up in any halo at time step 2 end up in halo 2; ii) 
more than 50\% of the particles in halo 2 come from halo 1.

\subsection{''Twin'' matching and orbital parameters}
\label{sec:tools:twins}

\begin{figure}
\psfig{figure=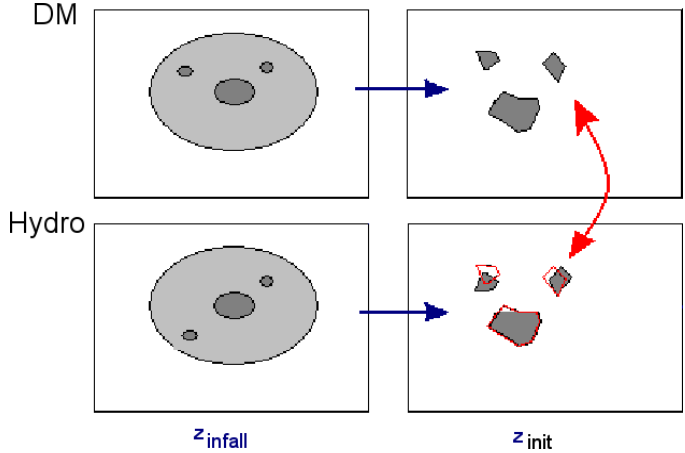,width=0.5\textwidth}
\caption{Illustration of satellite mapping scheme.}
\label{fig:tools:twins}
\end{figure}

Since we aim to compare the hydrodynamical and pure DM simulations by
studying satellite ''twins'', i.e. satellites that formed in both
simulations from the same perturbations at the initial time, we perform the following
steps to identify these satellite pairs:

\begin{enumerate}

\item We use the AHF halo finder to identify a given satellite in the
  DM simulation at the redshift $\bar z$ of interest.

\item We track the position of the satellite DM particles back in time to the initial condition file, in order to determine its original
  Lagrangian region.

\item We look for DM particles within the set of hydro initial conditions in
  the same Lagrangian region and map them forward in time to the same
  redshift $\bar z$.

\item Using the satellite catalogue for the hydro simulation we check
  in which satellite(s) these dm particles end up. We then use the
  same criteria adopted in the merger tree construction to
  determine the correct ``twin'' pair (see
  Fig. \ref{fig:tools:twins}).

\end{enumerate}

We have explicitly tested that the exact same results for the
satellite mapping follow starting either from the DM or the hydro
simulation. Overall we find more than 200 satellites pairs and 76 of
them survived down to $z=0$ in both simulations. We will refer to this
last subset of satellites as the twin population at redshift zero.

Finally we compute the orbital parameters (apo-center and peri-center
distances) for each pair of twin satellites. In order to do that we
integrate the satellite orbit in an effective static potential
parametrized using the \emph{power-law logarithmic slope} (PoLLS)
model \citep{cardone2005}:

\begin{equation}
\rho^\mathrm{PoLLS}(r) = \rho_{-2} \exp( -\frac 2 \gamma 
\left[ \left( \frac r {R_{-2}} \right)^\gamma -1 \right]) 
\label{eq:densPoLLS}
\end{equation}
\noindent
where $\rho_{-2}$ is the density at the $R_{-2}$ distance, which is where the profile slope
is equal to $-2$ (i.e. the singular isothermal sphere profile) and $\gamma$ is related
the central slope of the profile. All these parameters are fitted to
match the actual potential in either the DM or hydro simulation.  In
principle since we have constructed the merger trees it would have
been possible to compute the satellite orbital parameters directly
from the simulation.  On the other hand, given the time sampling of
our snapshots we could end up with over/under estimating the
peri/apo-centre distances. So we decided to use a frozen potential in
order to have a more reliable orbital parameter estimate.

\section{Results} 
\label{sec:res}

In this section we study how the presence of baryons affects the properties of
satellites in our simulations.  We include in our analysis only substructures
with at least  $M_{tot}> 10^8 \Msun$.  We combine together the results of G0, G1 and G2 (which
give similar results when analyzed individually) and mainly focus on ``twin''
haloes. This gives us a sample of more than 70 satellite couples at $z=0$.  In
the following we refer to the dissipationless simulations as pure DM or Nbody,
while we use the therm ``hydro'' for dissipational simulation that include gas
and stars.

\begin{figure}
\psfig{figure=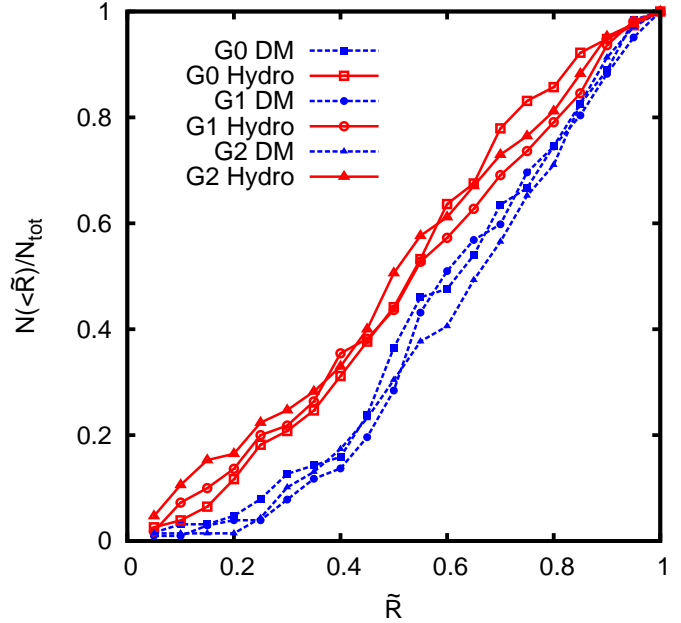,width=0.5\textwidth,angle=270}
\caption{\scriptsize Number density profile of ALL satellites at $z=0$.}
\label{fig:eval_sat:cumm_all}
\end{figure}

\begin{figure}
\psfig{figure=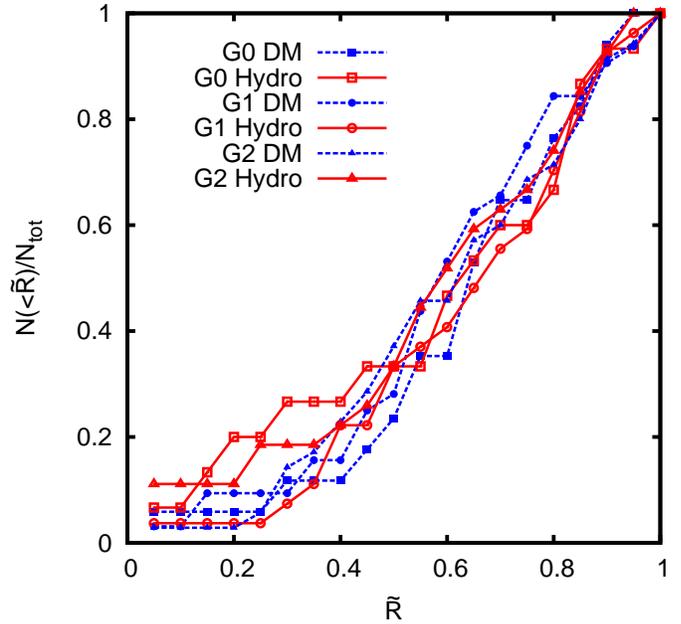,width=0.5\textwidth,angle=270}
\caption{\scriptsize Number density profile of TWIN satellites at $z=0$}.
\label{fig:eval_sat:cumm_twins}
\end{figure}

Figure \ref{fig:eval_sat:cumm_all} shows the cumulative radial
distribution (for all satellites) in our three galaxies in the two
different simulation runs.  As first noticed by \cite{maccio2006} and
later confirmed by other studies
\citep[e.g.][]{libeskind2010,romano2010} the radial distribution of
satellites is clearly more concentrated in hydro simulations (solid
red line) than in DM ones (dashed blue line).  This trend is partially
confirmed when we
restrict our analysis to twin satellites as shown in Figure
\ref{fig:eval_sat:cumm_twins}, at least in the inner region.
We should point out that ``twin''
satellites are a biased sub sample of the whole satellite population,
since, by construction, we restrict ourselves to those satellites that
survived until $z=0$ in both simulations. This biased selection
explains the difference in the radial distribution profiles of the whole population and of the twin
sample in the hydro and DM runs (especially in the external regions).


\begin{figure}
\psfig{figure=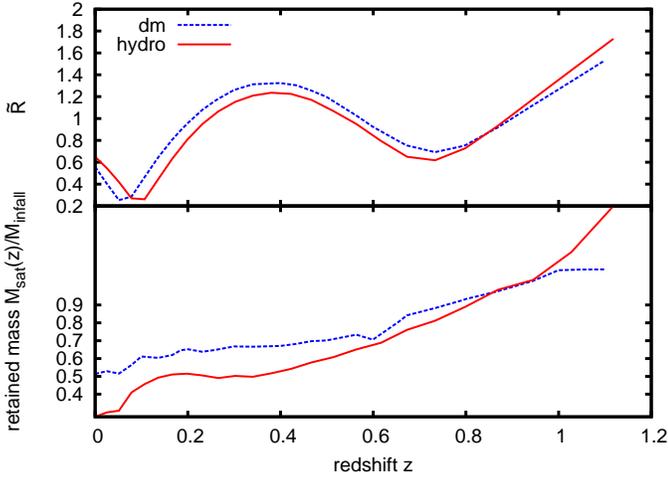,width=0.5\textwidth,angle=270}
\caption{\scriptsize Evolution of radial distance (upper panel) and mass (lower panel) 
over time for a satellite that lost more mass
in the hydrodynamical simulation than in the pure DM one.}.
\label{fig:plots:lessDMLoss}
\end{figure}

In order to understand the physical drivers behind the results of Fig.
\ref{fig:eval_sat:cumm_all} and Fig. \ref{fig:eval_sat:cumm_twins}, we
exam individual mass accretion and dynamical histories of all
twin couples from $z=1.2$ to the present time. 
This is the redshift at which the latest major merger happen (namely for the
G1 halo), after $z=1.2$ our galaxies have a quiet dynamical history that allows to
study in details the satellites properties and orbits
Figure \ref{fig:plots:lessDMLoss} shows the evolution with redshift of the
distance from the halo centre (upper panel) and the total mass (lower
panel) of an hydro subhalo (solid line) and its dm twin (dashed
curve).  The two satellites enter the virial radius of the main object
($\tilde R=1$) at the same time and have similar orbits, ending
at $z=0$ at roughly the same distance from the halo centre.  In this
particular case the mass loss is higher in the hydro case, where the
satellite is able to retain only 30\% of its mass (50\% in the pure DM
run).  This figure is similar to Figure 8 and 9 of
\cite{libeskind2010}, even though the role of DM and hydro are
reversed.

\begin{figure}
\psfig{figure=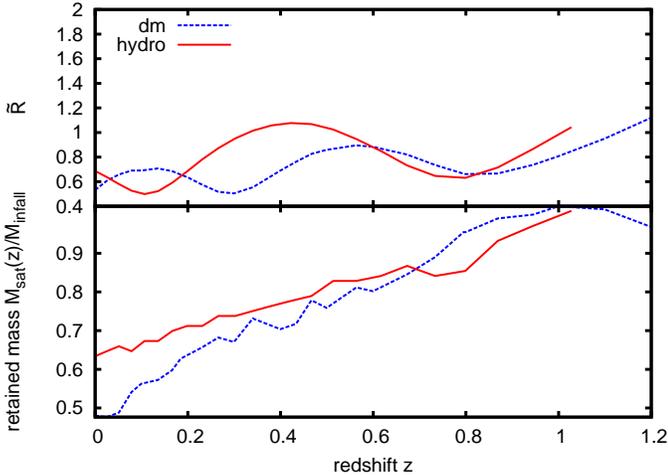,width=0.5\textwidth,angle=270}
\caption{\scriptsize Same as figure \ref{fig:plots:lessDMLoss}, but for a satellite that 
lost less mass in the hydrodynamical run than in the pure DM one.}
\label{fig:plots:moreDMLoss}
\end{figure}

The behaviour presented in
Fig. \ref{fig:plots:lessDMLoss} is rare, however.  The majority of satellites have more
complicated dynamical histories as shown in Fig.
\ref{fig:plots:moreDMLoss}. There, it is immediately noticeable that the
twins in this case do not enter the virial radius of the parent halo at the same
time. The infall time is $z_{infall}=0.93$ in the hydro case and
$z_{infall}=1.25$ in the pure DM one.  The orbits are moreover very
different: in the hydro simulation the satellite has a very large
apo-center that brings it outside the virial radius at $z\approx 0.5$
before being re-accreted; this does not happen in the pure DM run.  As a
consequence, the orbits are practically uncorrelated, and satellites experience
different mass loss rates.

\begin{figure}
\psfig{figure=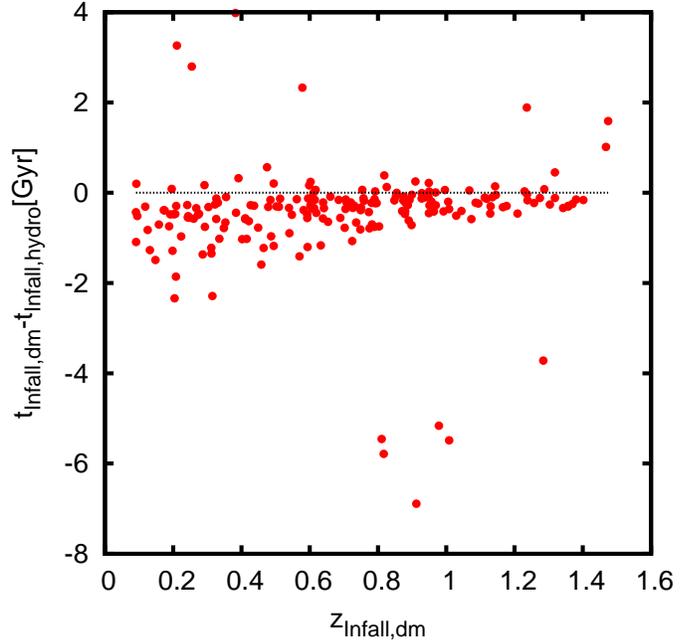,width=0.5\textwidth,angle=270}
\caption{\scriptsize Difference in infall time (e.g. crossing of the virial  radius) for twins. 
$t_{infall}$ is the time measured from the Big-Bang.}
\label{fig:plots:infall_times}
\end{figure}

An interesting feature revealed by Fig. \ref{fig:plots:moreDMLoss} is
the difference in the time of accretion between the hydro and DM runs. This is not
a peculiar behaviour of the selected satellite, but it is quite a
general trend, as shown in Fig. \ref{fig:plots:infall_times}.  For each
twin couple we compute the infall time (defined as the time at which
the satellite first crosses the virial radius of the main
object) in the hydro and pure DM runs and then we plot the difference
in these two times as a function of the infall redshift in the DM
case.

\begin{figure}
\psfig{figure=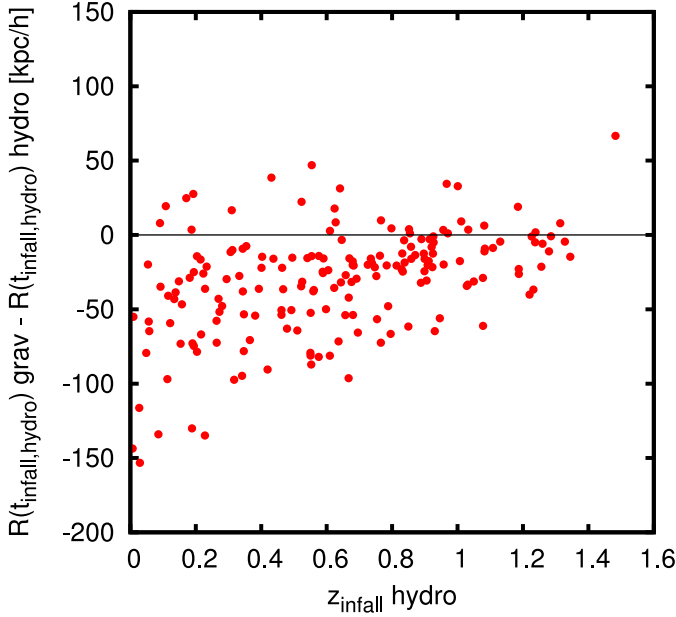,width=0.5\textwidth,angle=270}
\caption{\scriptsize Difference in physical distance from the center of the halo 
for twins at the time the hydro satellite crossed the virial radius.}
\label{fig:distacc}
\end{figure}

\begin{figure}
\psfig{figure=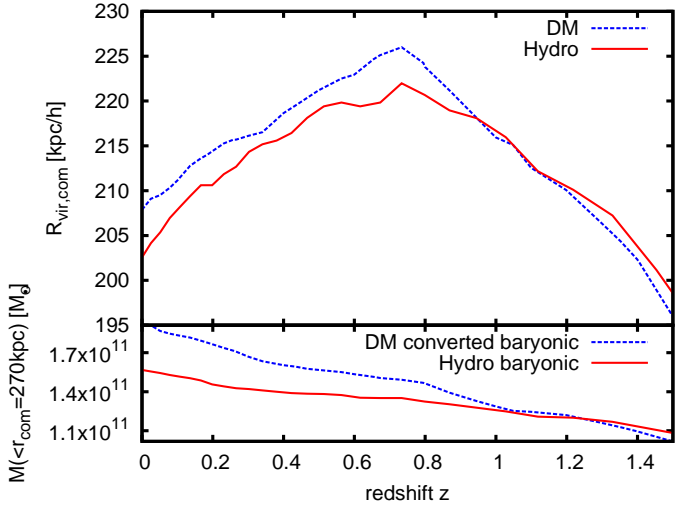,width=0.5\textwidth,angle=270}
\caption{\scriptsize Comparison of evolution of (top) comoving virial
  radius for G2 and (bottom) mass of (converted) baryonic matter
  inside sphere of 270 Mpc (comoving)}
\label{fig:eval_halo:Rvir}
\end{figure}

Most of the hydro satellites enter the virialized region later than
their pure DM counterparts, with an average delay of 0.7
Gyrs\footnote{Some haloes show a very large $\Delta t$. These are
  twins were one of the two subhaloes went in and out the virial
  radius before being accreted.  The so-called back splash haloes
  \citep{knebe10}}.  A possible explanation for this delayed accretion
might be time evolution in the main halo virial radius.  Fig.
\ref{fig:eval_halo:Rvir} shows $R_{vir}(z)$ for the hydro and pure DM
simulation of the G2 galaxy.  (We observe a similar effect for G0 and
G1 also.) In both runs the virial radius (in comoving coordinates)
grows until $z\approx 0.8$ and then decreases toward $z=0$.  Although
the DM run shows a larger value of $R_{vir}$ after $z=1$, this
difference is very small (about 5\%) and cannot explain the results of
Fig. \ref{fig:plots:infall_times}. This also confirm by figure \ref{fig:distacc}
where we plot the difference of physical distance from the center of the halo for twin
couples at the moment the hydro satellite crossed the virial radius. The plot
confirms that hydro satellites are substantially farther away compared to their 
DM counterparts at the time of accretion.

An alternative, possible, explanation arises from the point of view of the
initial conditions.  The hydro and pure DM runs differ only in that a
fraction $\Omega_b/\Omega_{dm}$ of the total mass in the hydro
simulation is represented by gas particles, while these gas particles
are converted to dark matter particles for the Nbody run. The lower
panel of Fig. \ref{fig:eval_halo:Rvir} shows the mass within a fixed
radius of 270 kpc due to baryonic particles (stars + gas) in the hydro
run and the same DM-converted particles in the pure DM run.  After
$z=1$ there, the amount of ``converted particles'' inside a sphere of
270 kpc exceeds the baryonic mass accumulated within the same region.
One possible explanation for this effect is that the pressure in the
baryonic hot gas slows down the accretion with respect to the
pressure-less ''converted'' dark matter component (especially
 in low density regions where the cooling time of the gas is
extremely long).  We speculate that the resulting delay in formation within the
hydro simulation in this case could explain the delayed accretion of
hydro satellites shown in Fig. \ref{fig:plots:infall_times}.
Nevertheless further studies are needed to confirm our hypothesis.

\subsection{Orbital Parameters}

The largely different time evolution of $\tilde R$ for twin haloes
shown in Fig. \ref{fig:plots:moreDMLoss} suggests that the radial
position at any fixed redshift is not well-suited for directly comparing
the orbits of the ``twin'' satellites.  The peri-centric and apo-centric
distances (see Section \ref{sec:tools:twins}), as well as the length of
the semi-major axis of the orbit $R_{avg} \equiv (R_{peri} + R_{apo})
/2.0$, seem a better choice.

\begin{figure}
\psfig{figure=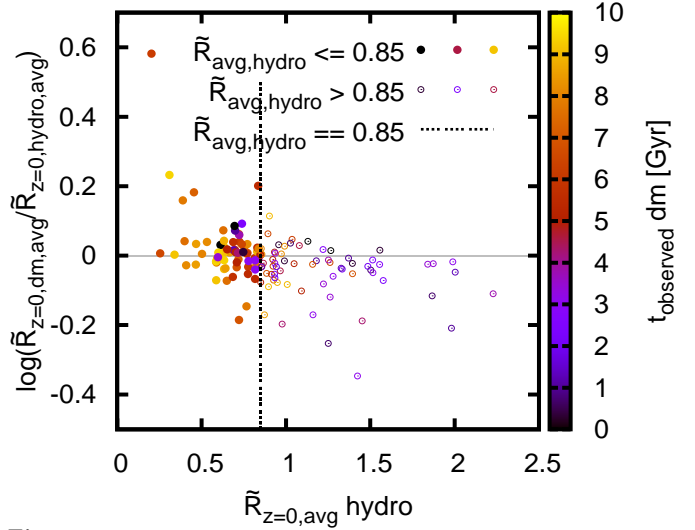,width=0.5\textwidth,angle=270}
\caption{\scriptsize Comparison of average distance of satellite
  ''twins'' at redshift $z=0$.  Each couple is color coded according
  to its DM observation time (i.e. the time at which they are closer
  than $2\times R_{vir}$ from the main halo center).}
\label{fig:plots:avgDist}
\end{figure}

Figure \ref{fig:plots:avgDist} shows the ratio of the average distance ($R_{avg}$) for 
hydro and pure DM satellites as a function of the average distance in the hydro run.
In the pure DM case $\tilde R_{avg} \equiv R_{avg}/R_{vir}$ tends to exceed the corresponding value 
in the hydrodynamical simulations for satellites that are close to the center. In 
the outer region of the halo the trend is reversed and 
satellites have a smaller average distance in the pure DM case.
This change in the radial behaviour seems to happen around $\tilde R_{avg} >0.85$. We will use this 
(somehow arbitrary) threshold to separate the inner and outer behaviour of our twin subhaloes in
the next figures.

In this same plot, twin couples are color coded according to their observation
time (i.e. the time at which they are closer than $2\times R_{vir}$ from the
main halo center). We find a very interesting correlation between the ratio of
$\tilde R_{avg}$ in the DM and hydro runs and this observation time.
Satellites with low $t_{obs}$ tend to live in the outskirts of the halo and
are on average closer to the halo center in the DM run than in the hydro case.
In contrast, ``old'' satellites (i.e. large $t_{obs}$) tend to live in the
inner region of the halo and are closer to the halo center in the hydro run
than in the pure DM.

\begin{figure}
\psfig{figure=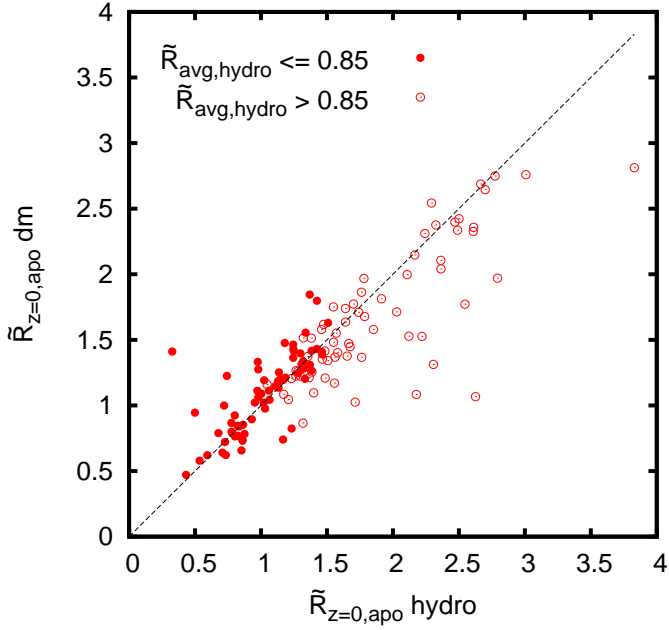,width=0.5\textwidth,angle=270}
\caption{\scriptsize Comparison of apo-centric distance of satellite ''twins'' at redshift $z=0$.
The apo-centric in computed using a frozen potential modelled according to eq: 1.}
\label{fig:plots:apoDist}
\end{figure}

This result can be understood in the following way: in the outer
region of the halo, dynamical friction is less important, and the
distance from the center of a given satellite mainly depends on its
accretion time. On the other hand, in the inner regions dynamical
friction is stronger in the hydro simulation (due to the central
stellar body), dragging satellites toward the center in a more
effective way compared to the pure DM case.
This picture is confirmed by the comparison between the apo-centre
distances for twins shown in Fig.  \ref{fig:plots:apoDist}. Satellites
in the outer region of the halo ($\tilde R_{avg,hydro}>0.85$) have a larger
apo-center distance in the hydro simulation, while the
situation is reversed for satellites further inside the halo.

\subsection{Mass loss}

\begin{figure}
\psfig{figure=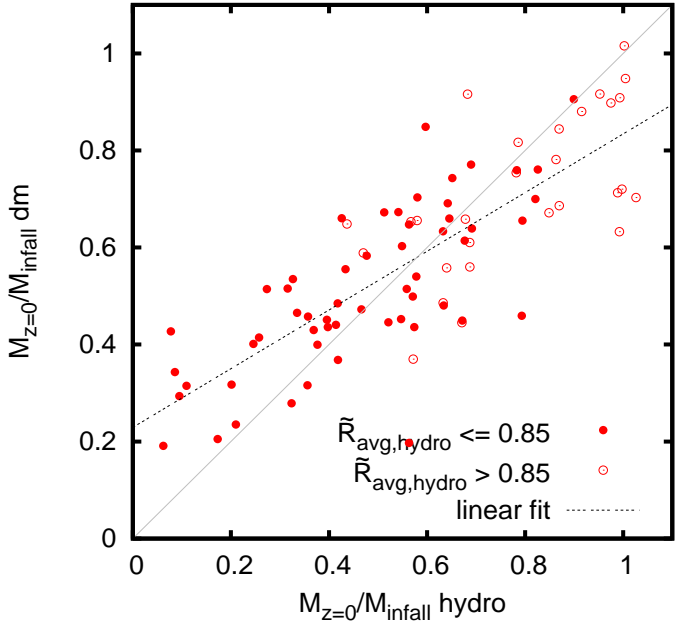,width=0.5\textwidth,angle=270}
\caption{\scriptsize 
Scatter plot of relative retained mass of twins in pure DM and hydrodynamical simulation (combined results for G0,G1,G2)}
\label{fig:eval_sat:MassLoss}
\end{figure}

The rate at which mass is removed from a given satellite depends on
the balance between the density profile of the satellite and the
density profile of the central halo.  \cite{libeskind2010} found that
satellites in hydro simulations experience a lower mass loss due to
their increased central density.  On the other hand \cite{romano2010}
found that the central region of hydro galaxies is depleted of
satellites at a faster rate than in pure DM simulations, pointing to a
faster and stronger mass loss in the hydro case.
Figure \ref{fig:eval_sat:MassLoss} shows a scatter plot of the
retained mass of all twins at $z=0$. Satellites that lose less than
60\% of their mass tend to retain more mass in the hydro simulation,
consistent with the findings of \cite{libeskind2010}.  But satellites that are heavily stripped tend to lose even more mass in the hydro simulation.  There is also a clear correlation between the
average position of the satellite and the amount of stripping.

\begin{figure}
\psfig{figure=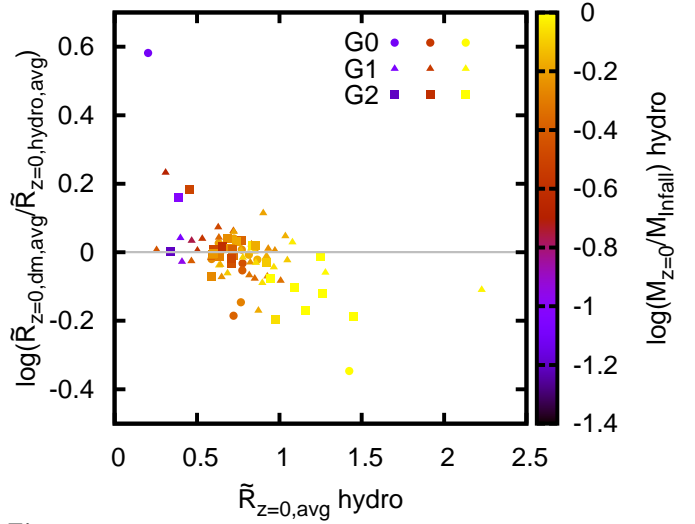,width=0.5\textwidth,angle=270}
\caption{\scriptsize Comparison of the ratio of the average distance in the DM and hydro sim as a 
function of the average distance in the hydro case. Twins are color coded according to their retained mass 
in the hydro simulation (combined results for G0,G1,G2). }
\label{fig:eval_sat:PeriApoMassLoss}
\end{figure}

This result becomes clearer when comparing the mass loss with the
orbital parameters of the satellite ($\tilde R_{avg}$), as in
Fig. \ref{fig:eval_sat:PeriApoMassLoss}. Satellites that live in the
external part of the halo tend to be at larger distances from the
center and retain more mass in the hydro simulation than in the pure
DM one. On the other hand, satellites in the central region are closer to
the center and more heavily stripped in the hydro simulation, pointing
to a more efficient dynamical friction and tidal stripping (in the
central regions) in the hydro simulation, in
agreement with the findings of \cite{romano2010}.

\begin{figure}
\psfig{figure=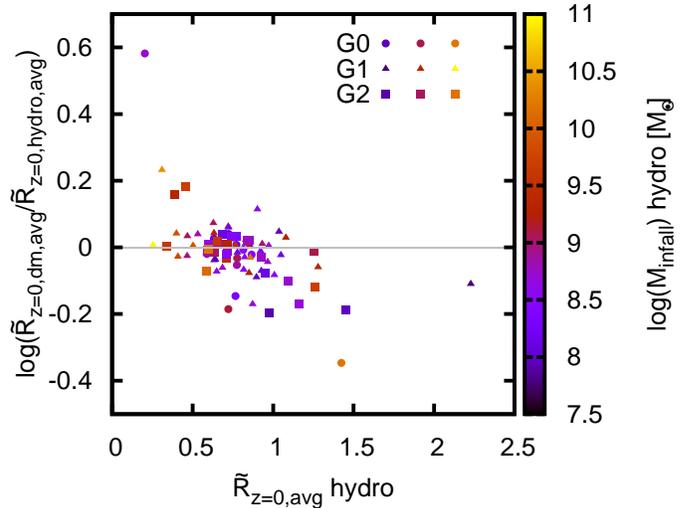,width=0.5\textwidth,angle=270}
\caption{\scriptsize Same as figure \ref{fig:eval_sat:PeriApoMassLoss} but with twins color coded
according to their mass at infall time in the hydro simulation (combined results for G0,G1,G2). }
\label{fig:eval_sat:PeriApoInfallMass}
\end{figure}

In Fig. \ref{fig:eval_sat:PeriApoInfallMass} satellites are
color-coded according to their total mass (in the hydro simulation) at
the time of infall. As expected from dynamical friction theory, more
massive satellites are living at $z=0$ closer to center than low mass
ones. This implies that satellites with large $M_{infall}$ will be even
closer to the center in hydro than in the DM simulations.

Together these results suggest a bimodal picture in
which high mass satellite at the time of accretion are {\it closer} to
the center and {\it more} heavily stripped in the hydro simulation
than in the pure DM, but the reverse for satellites with
low $M_{infall}$, which are on average {\it further} way from the center
and {\it less} stripped in the hydro simulation.

\subsection{Danger to the galaxy disk}

In this section we assess {\it if} and {\it how} the different orbital
parameters and mass loss rates found for hydro massive satellites
influence results previously obtained using pure Nbody simulations
\citep{font2001,read2008,villalobos2008,kazantzidis2009}.

\begin{figure}
\psfig{figure=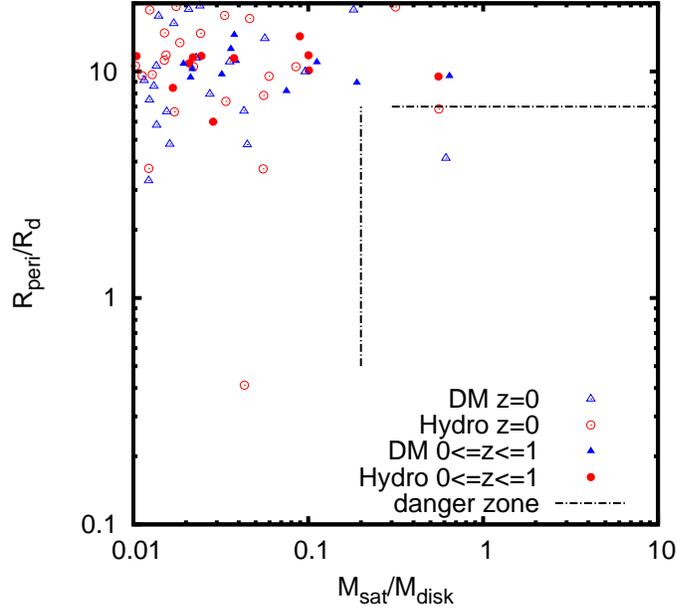,width=0.5\textwidth,angle=270}
\caption{\scriptsize Scatter plot of peri-centric distance versus
  satellite mass. Filled symbols thereby represent the satellites that
  cross within a radius of 50 kpc after $z=1$ while unfilled symbols
  mark the properties of surviving substructures at $z=0$ (twins
  only)}
\label{fig:kaz_twins}
\end{figure}

Figure \ref{fig:kaz_twins} shows a scatter plot of mass versus
peri-centric distance for two different twin substructure populations within
host galaxies G0-G2. It is equivalent to Fig. 1 of \cite{kazantzidis2009} and we have used the same values for the scaling quantities
$M_{disk}$ and $R_d$ as in their work, namely $M_{disk}=3.52 \times
10^{10} \Msun$ and $R_d=2.82$ kpc. 

The first substructure population in Fig. \ref{fig:kaz_twins} is comprised of all systems 
that have crossed an infall radius of $r_{\rm inf}=50\kpc$ from their 
host halo center since redshift $z=1$. This selection is empirically fixed 
to identify orbiting satellites that approach the central regions 
of the host potential and are thus likely to have a significant 
dynamical impact on the disk structure. We assign masses to the satellites
of this group at the simulation output time closest to the time of the first inward crossing 
of $r_{\rm inf}$, and then use the potential at this same simulation output to compute
the orbit peri-center. Note that a single distinct object of this population may be recorded 
multiple times, as one subhalo may undergo several passes through the 
central regions of its host with different masses and peri-centers. 
Many of these satellites suffer substantial mass loss prior to $z=0$.

The second subhalo population consists of all surviving
substructures at $z=0$.  The dotted line in Fig. \ref{fig:kaz_twins} encloses
an area in the $M_{\rm sub}- r_{\rm peri}$ plane corresponding to
satellites more massive than $0.2 M_{\rm disk}$ with peri-centers of
$r_{\rm peri} \lta 20\kpc$ ($r_{\rm peri} \lta 7 R_d$).  Subhaloes within this area should be effective perturbers, and following \cite{kazantzidis2009} we refer to this area as the ``danger
zone''.  We note, though, that we include this definition here for the purposes of illustration, only.

By this criterion, Figure~\ref{fig:kaz_twins} suggests that satellites in hydro simulation are
slightly less dangerous to disk
stability than their pure DM counterparts. This can be interpreted on the basis of results shown in the previous sections.
While the peri-centric distance does not substantially differ in hydro and DM runs,
hydrodynamical satellites face a stronger mass loss.  Since the danger for
galactic disks is higher the more massive the satellite, this
faster mass depletion for hydro satellites implies less dangerous perturbers for galactic disks.

\section{Discussion and conclusions}
\label{sec:conc}

With this work we aim to provide a detailed description of the
effects of baryonic physics on the properties of galactic satellites
and their evolution with redshift.  For this purpose we analyse three
different cosmological simulations of galaxy formation, each 
run twice: once as pure dark matter and another with the addition of gas
physics, including gas cooling, star formation and feedback.
For each run we create a comprehensive catalogue of the subhalo
population at $z=0$ which we trace back in time in order to study
mass loss, dynamical friction and evolution of satellite orbital
parameters.  Within this population we focus on a sub-sample of 
corresponding DM and hydro satellite pairs (``twins'') and their individual evolutionary histories.

Satellites are found to be more radially concentrated in the hydro
simulation than in the pure DM one, confirming earlier results by
\cite{maccio2006}.  This bias persists also for the twin population,
even if slightly less pronounced.  When we restrict our analysis to
the twin sub-sample we find that hydro satellites tend to enter the
virial radius of the parent halo later than the corresponding DM
subhaloes, with an average delay of 0.7 Gyrs.  This difference cannot
be ascribed to a difference in the evolution of $R_{vir}$ in the two
simulations, and we speculate that it is instead related to the
pressure support of the hot gas that acts against collapse, which, in
low density regions, is not counterbalanced by cooling. Nevertheless further studies 
are need to confirm our hypothesis.

Given the delay in the accretion time, the orbits of twin satellites
are often weakly correlated.  As a consequence, we find that the
radial position at a given redshift is not sufficient to describe the
satellite orbit. For this reason we define an average radius $R_{avg}$
for each satellite equal to the median between the apocenter and
pericenter distances of the satellite orbit, where these last two
quantities are obtained by integrating each satellite orbit in a fixed
potential resembling that of the halo.  We find that the ratio of the
average positions in the hydro and DM cases measured at redshift zero
correlates with the accretion time of the satellite and its
mass at that time. Moreover, we find that both the absolute mass loss
experienced by the satellite and the difference in mass loss in the
hydro and DM simulations also correlate with the subhalo average
distance at $z=0$.

We arrive at a final picture in which more massive satellites at the time
of accretion are {\it closer} to the center and {\it more} heavily
stripped in the hydro simulation than in the pure DM.  The situation
is reversed for satellites with low $M_{infall}$ that are on average
{\it further} way from the center and {\it less} stripped in the hydro
simulation.

This bimodality can be understood in the following way: in the outer
region of the halo, dynamical friction is less important, and the
distance from the center of a given satellite mainly depends on its
accretion time. Since hydro satellites tend to be accreted later, they
experience lower dynamical friction and mass stripping than their
pure DM counterparts.   In the inner regions, on the other hand, dynamical
friction is stronger in the hydro simulation (due to the central
stellar body) and drags satellites toward the center in a more
effective way compared to the pure DM case.  Although the
baryons provide a substantial glue to the subhaloes, the main halo
exhibits the same trend.  This assures a more efficient tidal stripping
of the hydro subhalo population, resulting in a larger mass loss.

During the making of this work, two other groups,
i.e. \cite{libeskind2010} and \cite{romano2010}, independently pursued
the same subject.  Both studies compared hydro and pure DM
simulations using an approach similar to ours, including a focus on
corresponding pairs of satellites.  The publication of Libeskind et
al. mainly considers the radial distribution of the satellites and a
comparison of the retained masses. Like us, they find a difference in the orbits of twins.  But in contrast to our
approach, \cite{libeskind2010} use the statistics of the radial orbit position, rather then, e.g. determine the orbital parameters and the resulting average distance as we do.  They find a larger
mass loss for the satellite in the pure DM simulation and interpret
this as a sign of the expected higher stability of the hydrodynamical
satellite, which our results are only able to confirm in the external region
of the halo.
\cite{romano2010} on the other hand, find an increased mass loss for the hydrodynamical satellite in the central region of the halo, in accordance with our
results.  They are also able to detect the final
disruption of a satellite, and have shown that the life expectancy of
the satellites in the hydrodynamical simulation is indeed shorter than
in the pure DM simulation, as our findings suggest.

We extend this result further in the last part of our study by investigating
the possible impact of different orbital parameters and mass loss in
hydrodynamical simulations, i.e. whether this translates into an increase, or
a reduction, in the danger that these satellites pose for the stability of a
possible stellar disk at the center of the parent halo \citep[e.g.][]{kazantzidis2009,moster2010}.  While the peri-centric distance
does not substantially differ in hydro and DM runs, hydrodynamical satellites
in the central region face a stronger mass loss. As the danger for galactic
disks is linked to the mass of the satellite, this faster mass depletion for
hydro satellites leads to less dangerous perturbers for galactic disks,
possibly easing the problem of the existence of thin disks in a Cold Dark
Matter Universe.

We emphasize, though, that, as correctly pointed out by
\cite{romano2010}, the effects of baryonic cooling in the center of
dark matter (sub)haloes can be altered (if not reversed) for a more
efficient feedback from stellar evolution and possibly central
super-massive black holes, which will expel baryons from the center and
decrease the central concentration of the prime halo
\citep[e.g.][]{elzant2004,mashchenko2006,governato2010}.

At this time, the direct comparison of our study of galactic satellites with observations is not possible. The Sloan Digital
Sky Survey, which greatly contributed to the discovery and study of
several Milky Way satellites \citep[e.g.][]{koposov2008,dejong2010},
covers only a single patch of the sky.   Future surveys
like Pan-Starrs, which will provide a more comprehensive map of the (northern)
sky, promise a better understanding of the properties and spatial
distribution of satellite galaxies orbiting around the Milky Way.

\section*{Acknowledgements} 

The authors are in debt with Sharon E. Meidt for her useful comments
on an earlier version of this paper and for revising the text of the
final version.  We also thank the referee of our paper, Noam
Libeskind, for his very constructive report, that helped in
substantially improve the presentation of our results.  Numerical
simulations were performed on the PIA and on PanStarrs2 clusters of
the Max-Planck-Institut f\"ur Astronomie at the Rechenzentrum in
Garching.  AVM thanks A. Knebe for his help with the {\sc AHF}
halofinder.


\bibliographystyle{mn2e}
\bibliography{literatur}

\begin{thebibliography}{}

\bibitem[\protect\citeauthoryear{{Bailin}, {Kawata}, {Gibson}, {Steinmetz},
  {Navarro}, {Brook}, {Gill}, {Ibata}, {Knebe}, {Lewis} \& {Okamoto}}{{Bailin}
  et~al.}{2005}]{bailin2005}
{Bailin} J.,  {Kawata} D.,  {Gibson} B.~K.,  {Steinmetz} M.,  {Navarro} J.~F.,
  {Brook} C.~B.,  {Gill} S.~P.~D.,  {Ibata} R.~A.,  {Knebe} A.,  {Lewis} G.~F.,
     {Okamoto} T.,  2005, ApJL, 627, L17

\bibitem[\protect\citeauthoryear{{Cardone}, {Piedipalumbo} \&
  {Tortora}}{{Cardone} et~al.}{2005}]{cardone2005}
{Cardone} V.~F.,  {Piedipalumbo} E.,    {Tortora} C.,  2005, MNRAS, 358, 1325

\bibitem[\protect\citeauthoryear{{de Jong}, {Martin}, {Rix}, {Smith}, {Jin} \&
  {Macci{\`o}}}{{de Jong} et~al.}{2010}]{dejong2010}
{de Jong} J.~T.~A.,  {Martin} N.~F.,  {Rix} H.,  {Smith} K.~W.,  {Jin} S.,
  {Macci{\`o}} A.~V.,  2010, \apj, 710, 1664

\bibitem[\protect\citeauthoryear{{De Lucia}, {Kauffmann}, {Springel}, {White},
  {Lanzoni}, {Stoehr}, {Tormen} \& {Yoshida}}{{De Lucia}
  et~al.}{2004}]{deLucia2004}
{De Lucia} G.,  {Kauffmann} G.,  {Springel} V.,  {White} S.~D.~M.,  {Lanzoni}
  B.,  {Stoehr} F.,  {Tormen} G.,    {Yoshida} N.,  2004, \mnras, 348, 333

\bibitem[\protect\citeauthoryear{{Diemand}, {Kuhlen} \& {Madau}}{{Diemand}
  et~al.}{2007}]{diemand2007}
{Diemand} J.,  {Kuhlen} M.,    {Madau} P.,  2007, \apj, 667, 859

\bibitem[\protect\citeauthoryear{{Duffy}, {Schaye}, {Kay}, {Dalla Vecchia},
  {Battye} \& {Booth}}{{Duffy} et~al.}{2010}]{duffy2010}
{Duffy} A.~R.,  {Schaye} J.,  {Kay} S.~T.,  {Dalla Vecchia} C.,  {Battye}
  R.~A.,    {Booth} C.~M.,  2010, MNRAS, 405, 2161

\bibitem[\protect\citeauthoryear{{El-Zant}, {Hoffman}, {Primack}, {Combes} \&
  {Shlosman}}{{El-Zant} et~al.}{2004}]{elzant2004}
{El-Zant} A.~A.,  {Hoffman} Y.,  {Primack} J.,  {Combes} F.,    {Shlosman} I.,
  2004, ApJL, 607, L75

\bibitem[\protect\citeauthoryear{{Font}, {Navarro}, {Stadel} \& {Quinn}}{{Font}
  et~al.}{2001}]{font2001}
{Font} A.~S.,  {Navarro} J.~F.,  {Stadel} J.,    {Quinn} T.,  2001, ApJL, 563,
  L1

\bibitem[\protect\citeauthoryear{{Gao}, {White}, {Jenkins}, {Stoehr} \&
  {Springel}}{{Gao} et~al.}{2004}]{gao2004}
{Gao} L.,  {White} S.~D.~M.,  {Jenkins} A.,  {Stoehr} F.,    {Springel} V.,
  2004, MNRAS, 355, 819

\bibitem[\protect\citeauthoryear{{Gnedin}, {Kravtsov}, {Klypin} \&
  {Nagai}}{{Gnedin} et~al.}{2004}]{gnedin2004}
{Gnedin} O.~Y.,  {Kravtsov} A.~V.,  {Klypin} A.~A.,    {Nagai} D.,  2004, ApJ,
  616, 16

\bibitem[\protect\citeauthoryear{{Governato}, {Brook}, {Mayer}, {Brooks},
  {Rhee}, {Wadsley}, {Jonsson}, {Willman}, {Stinson}, {Quinn} \&
  {Madau}}{{Governato} et~al.}{2010}]{governato2010}
{Governato} F.,  {Brook} C.,  {Mayer} L.,  {Brooks} A.,  {Rhee} G.,  {Wadsley}
  J.,  {Jonsson} P.,  {Willman} B.,  {Stinson} G.,  {Quinn} T.,    {Madau} P.,
  2010, Nature, 463, 203

\bibitem[\protect\citeauthoryear{{Governato}, {Willman}, {Mayer}, {Brooks},
  {Stinson}, {Valenzuela}, {Wadsley} \& {Quinn}}{{Governato}
  et~al.}{2007}]{Governato2007}
{Governato} F.,  {Willman} B.,  {Mayer} L.,  {Brooks} A.,  {Stinson} G.,
  {Valenzuela} O.,  {Wadsley} J.,    {Quinn} T.,  2007, MNRAS, 374, 1479

\bibitem[\protect\citeauthoryear{{Haardt} \& {Madau}}{{Haardt} \&
  {Madau}}{1996}]{Haardt1996}
{Haardt} F.,  {Madau} P.,  1996, APJ, 461, 20

\bibitem[\protect\citeauthoryear{{Katz}}{{Katz}}{1992}]{Katz1992}
{Katz} N.,  1992, ApJ, 391, 502

\bibitem[\protect\citeauthoryear{{Kazantzidis}, {Zentner}, {Kravtsov},
  {Bullock} \& {Debattista}}{{Kazantzidis} et~al.}{2009}]{kazantzidis2009}
{Kazantzidis} S.,  {Zentner} A.~R.,  {Kravtsov} A.~V.,  {Bullock} J.~S.,
  {Debattista} V.~P.,  2009, ApJ, 700, 1896

\bibitem[\protect\citeauthoryear{{Kennicutt}
  Jr.}{{Kennicutt}}{1998}]{kennicutt1998}
{Kennicutt} Jr. R.~C.,  1998, ApJ, 498, 541

\bibitem[\protect\citeauthoryear{{Klypin}, {Kravtsov}, {Valenzuela} \&
  {Prada}}{{Klypin} et~al.}{1999}]{klypin1999}
{Klypin} A.,  {Kravtsov} A.~V.,  {Valenzuela} O.,    {Prada} F.,  1999, \apj,
  522, 82

\bibitem[\protect\citeauthoryear{{Knebe}, {Libeskind}, {Knollmann},
  {Martinez-Vaquero}, {Yepes}, {Gottloeber} \& {Hoffman}}{{Knebe}
  et~al.}{2010}]{knebe10}
{Knebe} A.,  {Libeskind} N.~I.,  {Knollmann} S.~R.,  {Martinez-Vaquero} L.~A.,
  {Yepes} G.,  {Gottloeber} S.,    {Hoffman} Y.,  2010, ArXiv e-prints

\bibitem[\protect\citeauthoryear{{Knollmann} \& {Knebe}}{{Knollmann} \&
  {Knebe}}{2009}]{Knollmann2009}
{Knollmann} S.~R.,  {Knebe} A.,  2009, APJS, 182, 608

\bibitem[\protect\citeauthoryear{{Komatsu}, {Dunkley}, {Nolta}, {Bennett},
  {Gold}, {Hinshaw}, {Jarosik}, {Larson}, {Limon}, {Page}, {Spergel},
  {Halpern}, {Hill}, {Kogut}, {Meyer}, {Tucker}, {Weiland}, {Wollack} \&
  {Wright}}{{Komatsu} et~al.}{2009}]{komatsu2009}
{Komatsu} E.,  {Dunkley} J.,  {Nolta} M.~R.,  {Bennett} C.~L.,  {Gold} B.,
  {Hinshaw} G.,  {Jarosik} N.,  {Larson} D.,  {Limon} M.,  {Page} L.,
  {Spergel} D.~N.,  {Halpern} M.,  {Hill} R.~S.,  {Kogut} A.,  {Meyer} S.~S.,
  {Tucker} G.~S.,  {Weiland} J.~L.,  {Wollack} E.,    {Wright} E.~L.,  2009,
  \apjs, 180, 330

\bibitem[\protect\citeauthoryear{{Koposov}, {Belokurov}, {Evans}, {Hewett},
  {Irwin}, {Gilmore}, {Zucker}, {Rix}, {Fellhauer}, {Bell} \&
  {Glushkova}}{{Koposov} et~al.}{2008}]{koposov2008}
{Koposov} S.,  {Belokurov} V.,  {Evans} N.~W.,  {Hewett} P.~C.,  {Irwin} M.~J.,
   {Gilmore} G.,  {Zucker} D.~B.,  {Rix} H.,  {Fellhauer} M.,  {Bell} E.~F.,
  {Glushkova} E.~V.,  2008, \apj, 686, 279

\bibitem[\protect\citeauthoryear{{Kravtsov}, {Berlind}, {Wechsler}, {Klypin},
  {Gottl{\"o}ber}, {Allgood} \& {Primack}}{{Kravtsov}
  et~al.}{2004}]{kravtsov2004}
{Kravtsov} A.~V.,  {Berlind} A.~A.,  {Wechsler} R.~H.,  {Klypin} A.~A.,
  {Gottl{\"o}ber} S.,  {Allgood} B.,    {Primack} J.~R.,  2004, \apj, 609, 35

\bibitem[\protect\citeauthoryear{{Libeskind}, {Yepes}, {Knebe},
  {Gottl{\"o}ber}, {Hoffman} \& {Knollmann}}{{Libeskind}
  et~al.}{2010}]{libeskind2010}
{Libeskind} N.~I.,  {Yepes} G.,  {Knebe} A.,  {Gottl{\"o}ber} S.,  {Hoffman}
  Y.,    {Knollmann} S.~R.,  2010, MNRAS, 401, 1889

\bibitem[\protect\citeauthoryear{{Macci{\`o}}, {Kang}, {Fontanot},
  {Somerville}, {Koposov} \& {Monaco}}{{Macci{\`o}} et~al.}{2010}]{maccio2010}
{Macci{\`o}} A.~V.,  {Kang} X.,  {Fontanot} F.,  {Somerville} R.~S.,  {Koposov}
  S.,    {Monaco} P.,  2010, \mnras, 402, 1995

\bibitem[\protect\citeauthoryear{{Macci{\`o}}, {Moore}, {Stadel} \&
  {Diemand}}{{Macci{\`o}} et~al.}{2006}]{maccio2006}
{Macci{\`o}} A.~V.,  {Moore} B.,  {Stadel} J.,    {Diemand} J.,  2006, MNRAS,
  366, 1529

\bibitem[\protect\citeauthoryear{{Mashchenko}, {Couchman} \&
  {Wadsley}}{{Mashchenko} et~al.}{2006}]{mashchenko2006}
{Mashchenko} S.,  {Couchman} H.~M.~P.,    {Wadsley} J.,  2006, Nature, 442, 539

\bibitem[\protect\citeauthoryear{{Moore}, {Ghigna}, {Governato}, {Lake},
  {Quinn}, {Stadel} \& {Tozzi}}{{Moore} et~al.}{1999}]{moore1999}
{Moore} B.,  {Ghigna} S.,  {Governato} F.,  {Lake} G.,  {Quinn} T.,  {Stadel}
  J.,    {Tozzi} P.,  1999, \apjl, 524, L19

\bibitem[\protect\citeauthoryear{{Moster}, {Macci{\`o}}, {Somerville},
  {Johansson} \& {Naab}}{{Moster} et~al.}{2010}]{moster2010}
{Moster} B.~P.,  {Macci{\`o}} A.~V.,  {Somerville} R.~S.,  {Johansson} P.~H.,
   {Naab} T.,  2010, MNRAS, 403, 1009

\bibitem[\protect\citeauthoryear{{Nagai} \& {Kravtsov}}{{Nagai} \&
  {Kravtsov}}{2005}]{nagai2005}
{Nagai} D.,  {Kravtsov} A.~V.,  2005, ApJ, 618, 557

\bibitem[\protect\citeauthoryear{{Quinn}, {Hernquist} \& {Fullagar}}{{Quinn}
  et~al.}{1993}]{quinn1993}
{Quinn} P.~J.,  {Hernquist} L.,    {Fullagar} D.~P.,  1993, ApJ, 403, 74

\bibitem[\protect\citeauthoryear{{Read}, {Lake}, {Agertz} \&
  {Debattista}}{{Read} et~al.}{2008}]{read2008}
{Read} J.~I.,  {Lake} G.,  {Agertz} O.,    {Debattista} V.~P.,  2008, \mnras,
  389, 1041

\bibitem[\protect\citeauthoryear{{Reed}, {Governato}, {Quinn}, {Gardner},
  {Stadel} \& {Lake}}{{Reed} et~al.}{2005}]{reed2005}
{Reed} D.,  {Governato} F.,  {Quinn} T.,  {Gardner} J.,  {Stadel} J.,    {Lake}
  G.,  2005, MNRAS, 359, 1537

\bibitem[\protect\citeauthoryear{{Romano-D{\'{\i}}az}, {Shlosman}, {Heller} \&
  {Hoffman}}{{Romano-D{\'{\i}}az} et~al.}{2009}]{romano2009}
{Romano-D{\'{\i}}az} E.,  {Shlosman} I.,  {Heller} C.,    {Hoffman} Y.,  2009,
  \apj, 702, 1250

\bibitem[\protect\citeauthoryear{{Romano-D{\'{\i}}az}, {Shlosman}, {Heller} \&
  {Hoffman}}{{Romano-D{\'{\i}}az} et~al.}{2010}]{romano2010}
{Romano-D{\'{\i}}az} E.,  {Shlosman} I.,  {Heller} C.,    {Hoffman} Y.,  2010,
  ApJ, 716, 1095

\bibitem[\protect\citeauthoryear{{Sommer-Larsen} \& {Limousin}}{{Sommer-Larsen}
  \& {Limousin}}{2010}]{sommer2010}
{Sommer-Larsen} J.,  {Limousin} M.,  2010, MNRAS, p.~1332

\bibitem[\protect\citeauthoryear{{Springel}, {Wang}, {Vogelsberger}, {Ludlow},
  {Jenkins}, {Helmi}, {Navarro}, {Frenk} \& {White}}{{Springel}
  et~al.}{2008}]{springel2008}
{Springel} V.,  {Wang} J.,  {Vogelsberger} M.,  {Ludlow} A.,  {Jenkins} A.,
  {Helmi} A.,  {Navarro} J.~F.,  {Frenk} C.~S.,    {White} S.~D.~M.,  2008,
  MNRAS, 391, 1685

\bibitem[\protect\citeauthoryear{{Springel}, {Yoshida} \& {White}}{{Springel}
  et~al.}{2001}]{springel2001}
{Springel} V.,  {Yoshida} N.,    {White} S.~D.~M.,  2001, New Astronomy, 6, 79

\bibitem[\protect\citeauthoryear{{Stadel}}{{Stadel}}{2001}]{stadel2001}
{Stadel} J.,  2001, PhD thesis, University of Washington

\bibitem[\protect\citeauthoryear{{Stinson}, {Seth}, {Katz}, {Wadsley},
  {Governato} \& {Quinn}}{{Stinson} et~al.}{2006}]{Stinson2006}
{Stinson} G.,  {Seth} A.,  {Katz} N.,  {Wadsley} J.,  {Governato} F.,
  {Quinn} T.,  2006, MNRAS, 373, 1074

\bibitem[\protect\citeauthoryear{{Toth} \& {Ostriker}}{{Toth} \&
  {Ostriker}}{1992}]{toth1992}
{Toth} G.,  {Ostriker} J.~P.,  1992, ApJ, 389, 5

\bibitem[\protect\citeauthoryear{{Velazquez} \& {White}}{{Velazquez} \&
  {White}}{1999}]{velazquez1999}
{Velazquez} H.,  {White} S.~D.~M.,  1999, MNRAS, 304, 254

\bibitem[\protect\citeauthoryear{{Villalobos} \& {Helmi}}{{Villalobos} \&
  {Helmi}}{2008}]{villalobos2008}
{Villalobos} {\'A}.,  {Helmi} A.,  2008, MNRAS, 391, 1806

\bibitem[\protect\citeauthoryear{{Wadsley}, {Stadel} \& {Quinn}}{{Wadsley}
  et~al.}{2004}]{Wadsley2004}
{Wadsley} J.~W.,  {Stadel} J.,    {Quinn} T.,  2004, New Astronomy, 9, 137

\bibitem[\protect\citeauthoryear{{Weinberg}, {Colombi}, {Dav{\'e}} \&
  {Katz}}{{Weinberg} et~al.}{2008}]{weinberg2008}
{Weinberg} S.~H.,  {Colombi} S.,  {Dav{\'e}} R.,    {Katz} N.,  2008, \apj,
  678, 6

\bibitem[\protect\citeauthoryear{{White} \& {Rees}}{{White} \&
  {Rees}}{1978}]{white1978}
{White} S.~D.~M.,  {Rees} M.~J.,  1978, MNRAS, 183, 341

\bibitem[\protect\citeauthoryear{{Zentner}, {Kravtsov}, {Gnedin} \&
  {Klypin}}{{Zentner} et~al.}{2005}]{zentner2005}
{Zentner} A.~R.,  {Kravtsov} A.~V.,  {Gnedin} O.~Y.,    {Klypin} A.~A.,  2005,
  \apj, 629, 219

\end{thebibliography}

\end{document}